\documentclass[final,a4paper,]{aastex702}

\newcommand{\Spix}{S_{\rm pix}}
\newcommand{\Sscale}{S_{\rm scale}}
\newcommand{\Scrit}{S_{\rm crit}}
\newcommand{\fsim}{f_{\rm sim}}
\newcommand{\NHtwo}{N_{\rm H_2}}
\newcommand{\Icont}{I_{\rm cont}}
\newcommand{\ICO}{I_{\rm CO}}

\shorttitle{Scale-Vector Alignment}
\shortauthors{Zhao et al.}

\usepackage{etoolbox}

\newcommand{\titleherofigure}{%
  \par\vspace{10pt}%
  \begin{center}
    \includegraphics[width=0.96\textwidth]{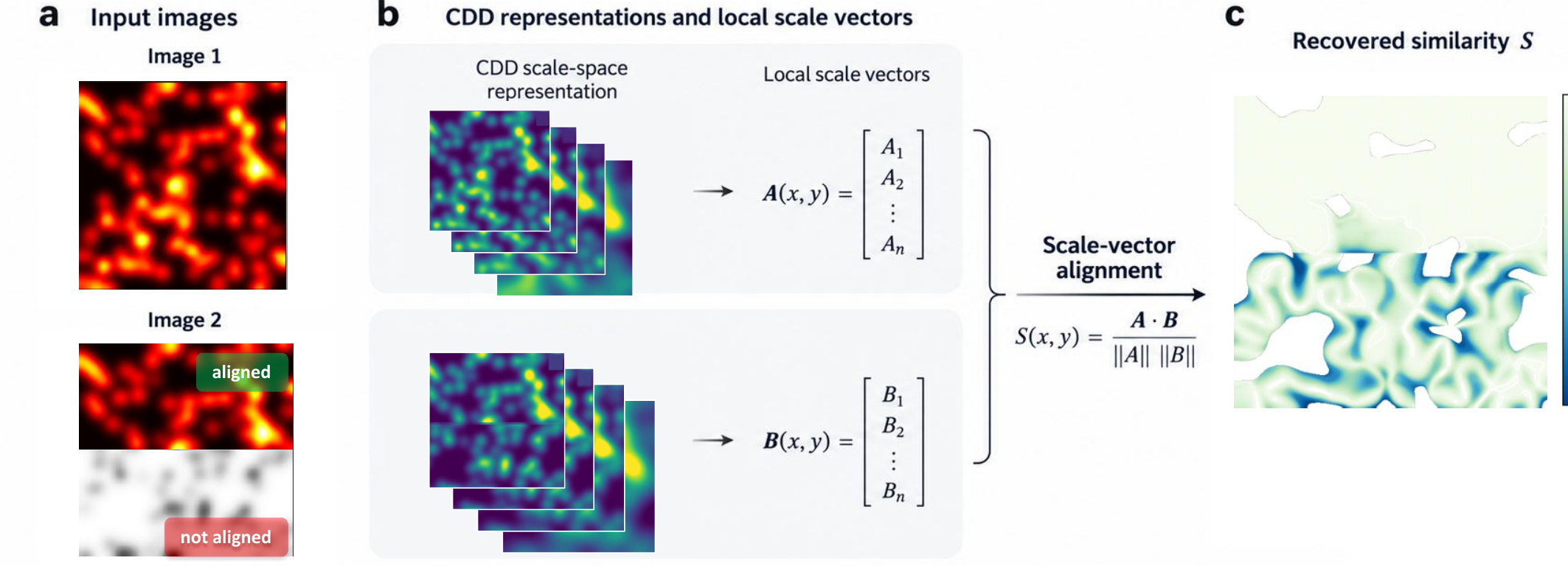}
  \end{center}
  \vspace{4pt}%
  \noindent\textit{Figure 1.} \textbf{Scale-vector alignment underlying the local CDD similarity.}
  The two synthetic input images are aligned in the upper half and contain different structure in the lower half. CDD maps each image into a scale-space representation, and the component amplitudes at the same position define local scale vectors $\mathbf{A}(x,y)$ and $\mathbf{B}(x,y)$. Their normalized inner product gives the local alignment. For visual simplicity, the pixel-wise similarity $\Spix(x,y)$ is denoted by $S(x,y)$ in the schematic. The recovered map gives $S\simeq1$ in the aligned region and lower values where the multiscale structure differs.
  \par\vspace{12pt}%
}

\makeatletter
\patchcmd{\titleblock@produce}
  {\frontmatter@abstract@produce}
  {\titleherofigure\frontmatter@abstract@produce}
  {}{\PackageError{SVA}{Could not insert title hero figure}{}}
\makeatother

\begin{document}

\title{Scale-Vector Alignment: A Scale-Aware Framework for Spatially Resolved Morphological Similarity in Astronomical Images}

\correspondingauthor{Guang-Xing Li,Keping Qiu}

\author[0000-0003-0596-6608]{Mengke Zhao}
\affiliation{School of Astronomy and Space Science, Nanjing University, 163 Xianlin Avenue, Nanjing 210023, Jiangsu, People's Republic of China}
\affiliation{Key Laboratory of Modern Astronomy and Astrophysics (Nanjing University), Ministry of Education, Nanjing 210023, Jiangsu, People's Republic of China}
\email{mkzhao628@gmail.com}

\author[0000-0003-3144-1952]{Guang-Xing Li}
\affiliation{South-Western Institute for Astronomy Research, Yunnan University, Kunming 650500, People's Republic of China}
\email[show]{gxli@ynu.edu.cn}

\author[0000-0002-5093-5088]{Keping Qiu}
\affiliation{School of Astronomy and Space Science, Nanjing University, 163 Xianlin Avenue, Nanjing 210023, Jiangsu, People's Republic of China}
\affiliation{Key Laboratory of Modern Astronomy and Astrophysics (Nanjing University), Ministry of Education, Nanjing 210023, Jiangsu, People's Republic of China}
\email[show]{kpqiu@nju.edu.cn}

\author[0000-0003-1275-5251]{Shanghuo Li}
\affiliation{School of Astronomy and Space Science, Nanjing University, 163 Xianlin Avenue, Nanjing 210023, Jiangsu, People's Republic of China}
\affiliation{Key Laboratory of Modern Astronomy and Astrophysics (Nanjing University), Ministry of Education, Nanjing 210023, Jiangsu, People's Republic of China}
\email{shli@nju.edu.cn}

\begin{abstract}
Astronomical maps made with different tracers are not expected to have identical morphology. Excitation, optical depth, chemistry, radiation, and ISM phase alter the response of a tracer, and the resulting differences can depend on both position and spatial scale. We propose scale-vector alignment, a scale-aware method based on Constrained Diffusion Decomposition (CDD). CDD decomposes an image into localized scale components; at each position, their amplitudes define a scale vector that describes how the measured intensity is distributed over spatial scale. We define the pixel-wise similarity $\Spix(x,y)$ as the normalized alignment of two local scale vectors. The normalization removes the overall amplitude, so $\Spix$ compares relative scale composition rather than absolute flux. We also define the scale-wise similarity $\Sscale(l)$ by comparing the two CDD component maps at each spatial scale. Spatial shifts are used to construct an empirical shifted reference distribution for $\Spix$. In Orion~A, the tracer with the highest similarity to the dust-derived column-density map changes from $^{12}$CO to $^{13}$CO to C$^{18}$O toward higher column density. In NGC~6334I(N), the line--continuum similarity decreases locally around the brightest compact structures, where radiative-transfer effects can alter the observed line morphology. In NGC~3627, CO is most similar to 21~$\mu$m emission, and $\Sscale$ reaches its maximum at an intermediate sub-kpc scale. The method measures where two tracers have similar multiscale structure and at which scales their spatial distributions agree. The implementation is publicly available at \url{https://github.com/meng-ke/Scale-Vector-Alignment}.
\end{abstract}

\section{Introduction}\label{sec:intro}

The interstellar medium (ISM) contains structure over a broad range of spatial scales \citep{2007ARA&A..45..565M,2012ARA&A..50...29C,2012A&ARv..20...55H,2025ARA&A..63....1B}. Turbulence, self-gravity, magnetic fields, radiation, and stellar feedback redistribute gas from cloud to core scales \citep{1987ARA&A..25...23S,2007ARA&A..45..565M,2012ARA&A..50...29C,2012A&ARv..20...55H,2025ARA&A..63....1B}. The intensity measured with a particular tracer is further set by abundance, excitation, optical depth, temperature, shielding, and radiative transfer \citep{2009ARA&A..47..427H,2013RvMP...85.1021T,2020ARA&A..58..727J,2022ARA&A..60..247W}. A tracer map is a physical response to the underlying medium rather than a direct map of its mass distribution. Different tracers can emphasize different layers or phases of the same region; this property has, for example, been used to trace magnetic-field structure with molecular lines of different optical depths \citep{2019ApJ...884..137H,2022ApJ...934...45Z}.

Tracer morphology depends on the physical conditions that produce the emission \citep{2007ARA&A..45..339B,2012ARA&A..50..531K,2013ARA&A..51..207B,2022ARA&A..60..247W}. Two tracers may follow the same structure in one part of a source but not in another. They may agree on cloud scales while separating on core scales, or the reverse. The comparison should retain both position and spatial scale.

CDD provides such a representation for positive intensity maps \citep{2022ApJS..259...59L}. Wavelet and related methods can measure scale-dependent structure and correlation, including wavelet cross-correlation and its weighted extension \citep{2001MNRAS.327.1145F,2016A&A...585A..98A}, but band-limited filtering can produce ringing and negative responses near sharp positive structures \citep{2022ApJS..259...59L}. CDD instead follows the image through a constrained diffusion sequence and obtains localized scale components from neighboring states of this sequence. For a positive input map, the components remain positive. Their amplitudes represent the contribution of the measured intensity over an ordered set of spatial scales. At a given pixel these amplitudes form a local scale vector. CDD scale representations have been used to estimate molecular-cloud widths and volume densities and to describe scale-dependent interferometric recovery \citep{2025arXiv250917369L,2026ApJ...997..345Z,2026arXiv260712396M}.

In this paper we use the CDD scale representation to compare two intensity fields. At each sky position, we compare the two local scale vectors and define the pixel-wise similarity $\Spix(x,y)$. At each spatial scale, we compare the corresponding CDD component maps and define the scale-wise similarity $\Sscale(l)$. The first quantity shows where the relative scale distributions agree; the second shows the scales on which the spatial distributions agree. We use spatial shifts to calibrate the local statistic and then examine how $\Spix$ changes with physical environment. Section~\ref{sec:method} defines the method. Section~\ref{sec:results} applies it to Orion~A, NGC~6334I(N), and NGC~3627. Synthetic tests and the input data products are given in the appendices.

\begin{figure*}[!b]
\centering
\includegraphics[width=0.95\textwidth]{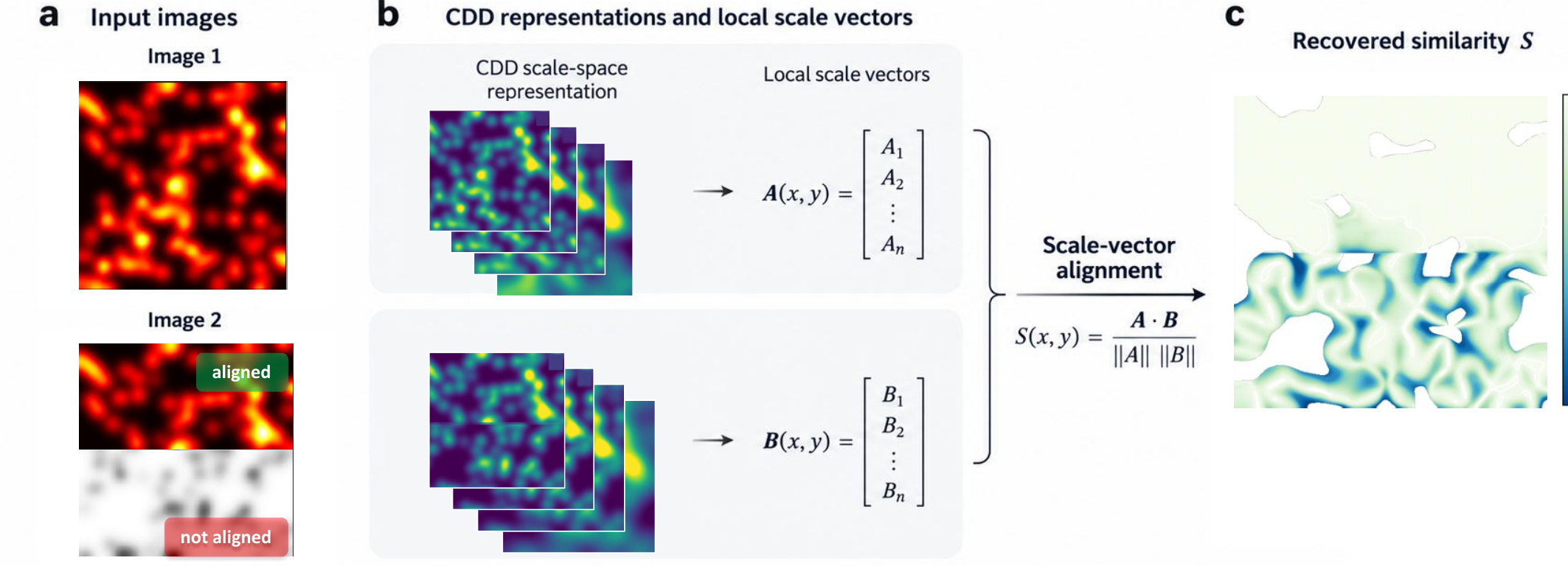}
\caption{Scale-vector alignment underlying the local CDD similarity. The two synthetic input images are aligned in the upper half and contain different structure in the lower half. CDD maps each image into a scale-space representation, and the component amplitudes at the same position define local scale vectors $\mathbf{A}(x,y)$ and $\mathbf{B}(x,y)$. Their normalized inner product gives the local alignment. For visual simplicity, the pixel-wise similarity $\Spix(x,y)$ is denoted by $S(x,y)$ in the schematic. The recovered map gives $S\simeq1$ in the aligned region and lower values where the multiscale structure differs.}
\label{fig:method}
\end{figure*}

\section{Method}\label{sec:method}

\subsection{Overview of procedure}\label{sec:procedure}

For two intensity fields $I_A(x,y)$ and $I_B(x,y)$ on the same pixel grid and at the same effective angular resolution, the calculation consists of the following steps:
\begin{enumerate}
\item Decompose both images with CDD on the same set of spatial scales, obtaining the component cubes $\{A_n(x,y)\}$ and $\{B_n(x,y)\}$.
\item At each valid position, form the local scale vectors $\mathbf{A}(x,y)$ and $\mathbf{B}(x,y)$ and compute their normalized alignment $\Spix(x,y)$.
\item Spatially shift the second CDD cube to construct an empirical reference distribution and determine the high-similarity threshold $\Scrit$.
\item At each CDD scale $l_n$, compare the corresponding component maps to obtain the scale-wise similarity $\Sscale(l_n)$.
\item When a physical coordinate $X$ is available, project the local similarity against $X$ and summarize the above-threshold correspondence with $\fsim(X)$.
\end{enumerate}
The quantities used in these steps are defined below.

\subsection{CDD as a local scale-space representation}\label{sec:cdd}

Let two images, $I_A(x,y)$ and $I_B(x,y)$, be defined on the same sky projection, pixel grid, and effective angular resolution. We decompose both images with CDD using the same set of scales. The $N$ component amplitudes at each position define
\begin{equation}
\mathbf{A}(x,y)=\left[A_1(x,y),A_2(x,y),\ldots,A_N(x,y)\right],
\end{equation}
\begin{equation}
\mathbf{B}(x,y)=\left[B_1(x,y),B_2(x,y),\ldots,B_N(x,y)\right],
\end{equation}
where the component index increases with spatial scale. We use the constrained CDD implementation of \citet{2022ApJS..259...59L}. The elements of each vector give the contribution of the local intensity to the sampled CDD scales.

The vector norm carries the overall amplitude, while the direction gives the relative distribution over scale. For tracers with different flux normalizations, the direction is the quantity of interest. We compare the vector directions. A common multiplicative rescaling of either tracer leaves the similarity unchanged. Figure~\ref{fig:method} illustrates this construction for two synthetic images.

The scale distribution must describe intrinsic spatial structure. In a field dominated by isolated unresolved point sources, the local scale vector can instead be set mainly by the instrumental point-spread function. Such fields are not the main use considered here. The same construction applies to other matched non-negative fields with resolved multiscale structure.

\subsection{Two complementary views of morphological similarity}\label{sec:similarity}

At each sky position, we compare the two local scale vectors through their normalized inner product,
\begin{equation}\label{eq:spix}
\Spix(x,y)=
\frac{\mathbf{A}(x,y)\cdot\mathbf{B}(x,y)}
{\left\|\mathbf{A}(x,y)\right\|
 \left\|\mathbf{B}(x,y)\right\|}
=
\frac{\displaystyle\sum_{n=1}^{N}A_n(x,y)B_n(x,y)}
{\displaystyle\sqrt{\sum_{n=1}^{N}A_n^2(x,y)}
 \sqrt{\sum_{n=1}^{N}B_n^2(x,y)}}.
\end{equation}
For the non-negative CDD components used here, $0\leq\Spix\leq1$. A value near unity means that the two tracers place the same relative fractions of their local intensity into the sampled scales. Because the vector norms are removed, $\Spix$ is insensitive to the absolute flux normalization.

At each spatial scale $l_n$, we instead compare the corresponding CDD component maps. Treating the valid pixels of $A_n(x,y)$ and $B_n(x,y)$ as spatial vectors gives
\begin{equation}\label{eq:sscale}
\Sscale(l_n)=
\frac{\displaystyle\sum_{x,y}A_n(x,y)B_n(x,y)}
{\displaystyle\sqrt{\sum_{x,y}A_n^2(x,y)}
 \sqrt{\sum_{x,y}B_n^2(x,y)}}.
\end{equation}
The curve $\Sscale(l)$ gives the similarity of the two spatial distributions as a function of scale. Its maximum identifies the CDD scale at which the two component maps have the closest spatial correspondence.

The two quantities use the same normalized inner product along different dimensions of the CDD representation. $\Spix(x,y)$ compares the distribution over spatial scale at each position, whereas $\Sscale(l)$ compares the spatial distribution at each scale. Their numerical values describe different properties and are not interchangeable.

\subsection{Empirical calibration of local similarity}\label{sec:null}

A large value of $\Spix$ does not by itself establish a physical correspondence, because structured images can align by chance. We construct an empirical reference by destroying the original positional correspondence while preserving the internal structure of each image. We generate 200 realizations by shifting the second CDD cube without wrap-around. Integer offsets are drawn over $\pm$ half the image size in each axis, with a minimum displacement of 30 pixels. The validity mask of the second image is shifted by the same offset, so a shifted similarity is retained only where the original analysis footprint and the shifted valid footprint overlap. The valid $\Spix$ values from all 200 realizations are pooled into a shifted reference sample $\{S^{\rm shift}_j\}_{j=1}^{N_0}$, where $N_0$ is the total number of pooled pixel values rather than the number of shifts. The shift preserves the scale structure within each map but removes the original pixel-by-pixel pairing. The 30-pixel exclusion is used to suppress residual local correspondence from small shifts. The empirical cumulative distribution is
\begin{equation}
\widehat{F}_0(S)=
\frac{1}{N_0}
\sum_{j=1}^{N_0}
\mathbb{I}\!\left(S^{\rm shift}_j\leq S\right),
\end{equation}
where $\mathbb{I}$ is the indicator function.

We define
\begin{equation}\label{eq:scrit}
\Scrit=\widehat{F}_0^{-1}(0.95),
\end{equation}
so that 5\% of the pooled shifted-reference values exceed $\Scrit$. We use $\Scrit$ as an empirical high-similarity threshold: pixels with $\Spix>\Scrit$ are classified as highly similar relative to the shifted reference distribution. Because the pooled values are spatially correlated, this classification is not interpreted as a formal per-pixel 95\% confidence level or $p<0.05$ significance test. $\Scrit$ is defined for the local statistic $\Spix$; $\Sscale$ is interpreted directly as a scale-dependent similarity.

To relate morphology to physical environment, we bin the local similarity by a coordinate $X$. In each bin we compute the median $\Spix$, its 16th--84th percentile interval, and the fraction
\begin{equation}\label{eq:fsim}
\fsim(X)=
\frac{N\left[\Spix(x,y)>\Scrit\right]}
{N_{\rm valid}},
\end{equation}
where both counts refer to valid pixels in the same bin. In the applications below, $X$ is $\NHtwo$, $\Icont$, or $\ICO$. In this form, $\Spix(x,y)$ retains the spatial information, $\fsim(X)$ gives the fraction of locations classified as highly similar relative to the shifted reference in a given physical regime, and $\Sscale(l)$ gives the scale dependence.

Details of image matching, masks, numerical precision, and scale-quality filtering are given in Appendix~\ref{app:implementation}. Figure~\ref{fig:method} illustrates the spatial-localization concept; independent synthetic tests of spatial localization and scale recovery are given in Appendix~\ref{app:synthetic}.

\section{Results and Discussion}\label{sec:results}

\subsection{Orion A: similarity across column-density regimes}\label{sec:orion}

Orion~A spans the column-density range over which the three CO isotopologues change their response to molecular-cloud structure. At a distance of about 400~pc, the cloud contains OMC-1, OMC-2/3, and L1641 and has been mapped extensively in molecular lines \citep{2019PASJ...71S...3N,2021PASJ...73S.239L}. We compare $^{12}$CO(1--0), $^{13}$CO(1--0), and C$^{18}$O(1--0) integrated-intensity maps from the CARMA--NRO Orion Survey \citep{2018ApJS..236...25K} with an H$_2$ column-density map derived from Herschel/Planck dust emission \citep{2014A&A...566A..45L}. The dust map provides a common reference for asking how the CO morphology changes with column density. The input maps and full-field scalar similarity maps are shown in Appendix~\ref{app:oriondata}.

\begin{figure*}[t]
\centering
\includegraphics[width=0.98\textwidth]{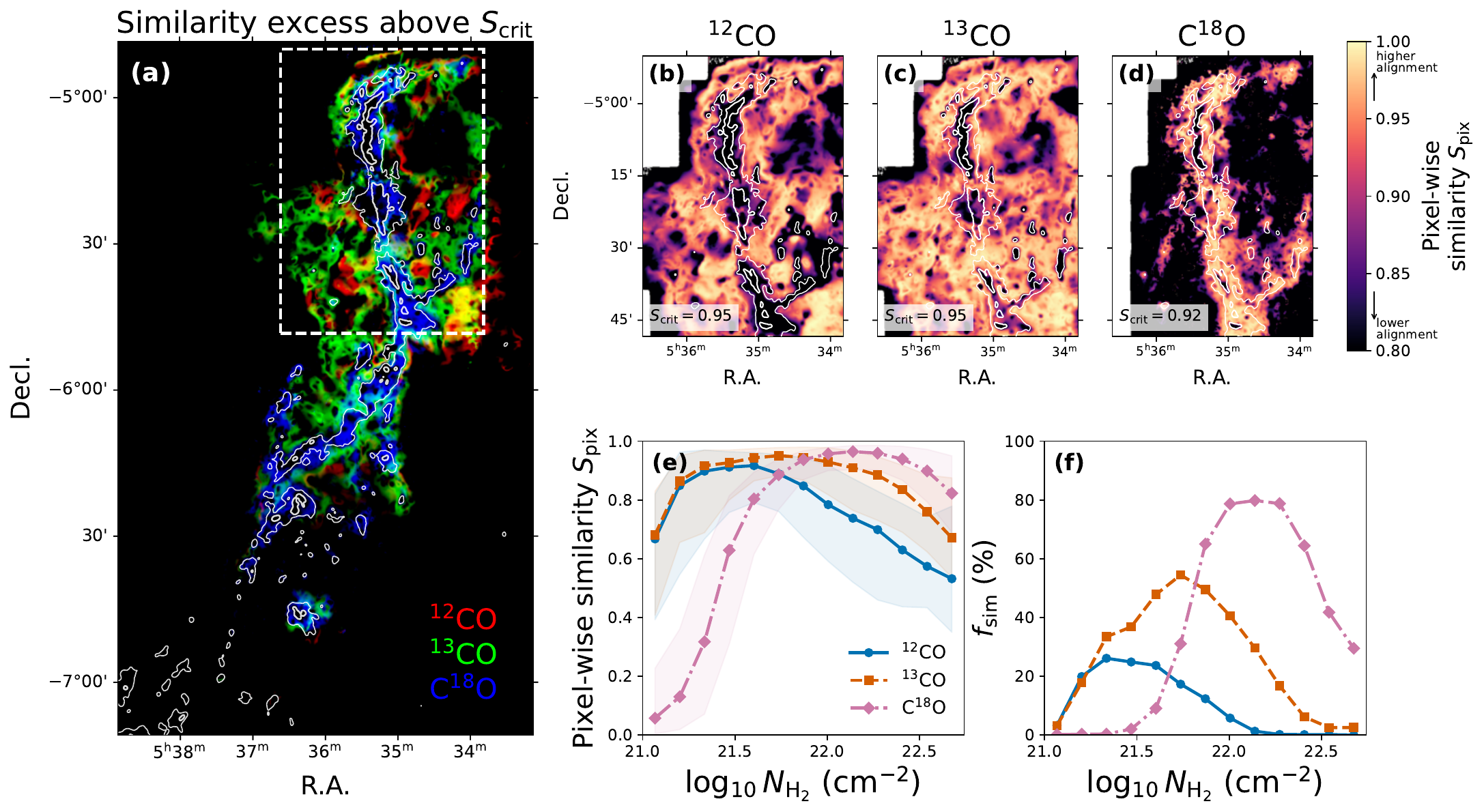}
\caption{Scale-vector similarity in Orion~A. 
(a) RGB composite of the normalized similarity excess,
$Z_i=\mathrm{clip}[(S_i-S_{{\rm crit},i})/(1-S_{{\rm crit},i}),0,1]$. For each tracer, the corresponding RGB channel is set to zero where $\Spix\leq\Scrit$, while increasing channel brightness indicates a larger similarity excess above $\Scrit$. Red, green, and blue represent $^{12}$CO, $^{13}$CO, and C$^{18}$O, respectively; mixed colors indicate locations where more than one tracer exceeds its own empirical reference threshold. The quantitative analysis uses the raw $\Spix$ values.
White contours mark $N_{\rm H_2}=10^{22}$ and $10^{22.5}\,{\rm cm}^{-2}$, and the dashed box marks the OMC-1--4 region. (b)--(d) Zoomed raw $\Spix$ maps in the boxed region; each panel gives its own $\Scrit$. (e) Median $\Spix$ as a function of $\NHtwo$, with 16th--84th percentile intervals. (f) Fraction of valid pixels above $\Scrit$. The column-density regime of above-threshold correspondence shifts from $^{12}$CO to $^{13}$CO to C$^{18}$O. Full-field scalar $\Spix$ maps are shown in Figure~\ref{fig:orionsimfull} in Appendix~\ref{app:oriondata}.}
\label{fig:orion}
\end{figure*}

The CO isotopologue with the highest similarity changes with column density. Figure~\ref{fig:orion}(a) compresses the three comparisons into an RGB map of similarity above the tracer-specific reference thresholds: different isotopologues exceed their own shifted reference in different parts of the cloud. The OMC-1--4 zooms in panels (b)--(d) resolve the dense northern region, where $^{12}$CO has lower similarity to the dust-derived column-density structure while $^{13}$CO and C$^{18}$O have higher similarity along dense structure. The binned trends in panels (e) and (f) show the same progression: $^{12}$CO reaches high similarity at lower $\NHtwo$ and declines toward the highest columns, $^{13}$CO remains similar over a broader intermediate range, and C$^{18}$O has the highest similarity at high $\NHtwo$. The $\fsim$ curves give the fraction of pixels that exceed the empirical reference threshold in each column-density regime.

The OMC-1 mismatch is consistent with the known radiative-transfer and chemical behavior of the CO isotopologues. Multiscale density reconstruction places dense OMC-1 gas at $n({\rm H_2})\sim10^{4}$--$10^{6}\,{\rm cm}^{-3}$ \citep{2026ApJ...997..345Z}. $^{12}$CO(1--0) is optically thick and can emphasize warm surfaces, velocity structure, outflows, and feedback rather than total column density \citep{2013ARA&A..51..207B}. The rarer isotopologues remain more sensitive to dense material, although their morphology is also affected by chemistry: the $^{13}$CO/C$^{18}$O abundance ratio varies with far-UV irradiation \citep{2019PASJ...71S...9I}, and C$^{18}$O(1--0) is largely optically thin in Orion~A but shows depletion toward the highest column densities \citep{2021PASJ...73..487T}. The change in $\Spix$ with $\NHtwo$ follows the regimes in which opacity, excitation, chemistry, and feedback alter the line emission relative to the dust-derived mass distribution.

\subsection{NGC 6334I(N): line--continuum coupling in a fragmenting protostellar system}\label{sec:ngc6334}

NGC~6334I(N) provides spatially resolved dust-continuum and molecular-line views of a fragmenting high-mass protostellar system. High-resolution observations resolve a population of compact millimeter sources \citep{2014ApJ...788..187H}, and recent ALMA data reveal a multiple protostellar system embedded in a Keplerian disk whose instability is consistent with disk fragmentation \citep{2025NatAs...9.1833L}. We use the Band~6 continuum and moment-0 maps of CH$_3$OH $4_{2,2}-3_{1,2}$, H$_2$CO $3_{03}-2_{02}$, and H$_2$CO $3_{22}-2_{21}$ from the data set analyzed by \citet{2025NatAs...9.1833L} (programs 2021.1.00713.S and 2022.1.00671.S). The comparison asks where line emission follows the compact dust structure and where excitation or radiative transfer changes that correspondence. The input images are shown in Appendix~\ref{app:ngc6334data}.

\begin{figure*}[t]
\centering
\includegraphics[width=0.96\textwidth]{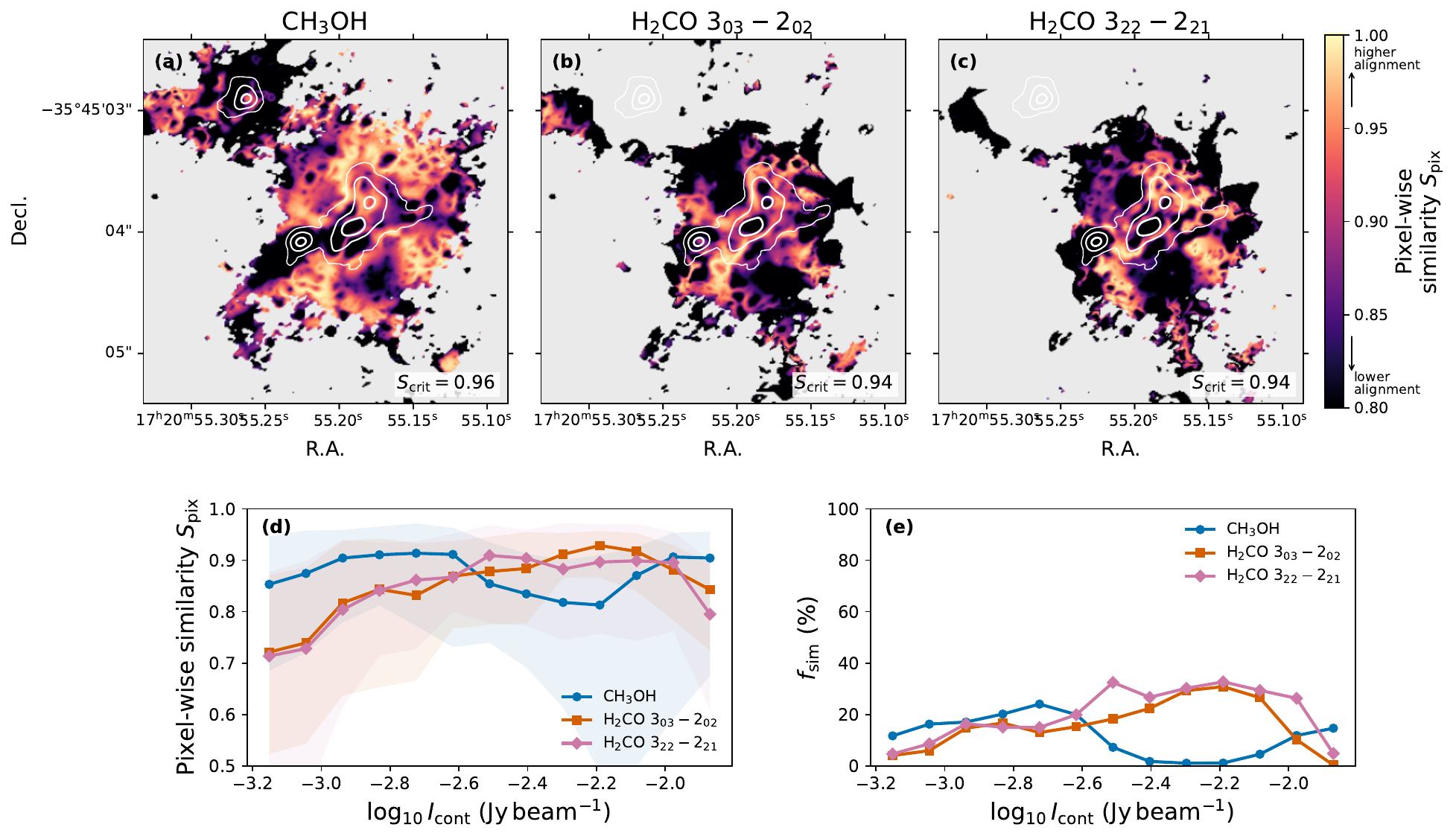}
\caption{Scale-vector similarity toward NGC~6334I(N). Top: pixel-wise similarity between the continuum and CH$_3$OH $4_{2,2}-3_{1,2}$, H$_2$CO $3_{03}-2_{02}$, and H$_2$CO $3_{22}-2_{21}$ moment-0 maps. Similarity is shown for pixels with continuum intensity above $5\sigma_{\rm cont}$, where $\sigma_{\rm cont}=1.5\times10^{-4}~{\rm Jy\,beam^{-1}}$; white contours mark 20, 50, and $110\sigma_{\rm cont}$. Bottom left: median $\Spix$ versus continuum intensity with 16th--84th percentile intervals. Bottom right: fraction of valid pixels above $\Scrit$.}
\label{fig:ngc6334}
\end{figure*}

The line--continuum similarity is not uniform across the compact source. In Figure~\ref{fig:ngc6334}, CH$_3$OH has the largest spatial extent of high $\Spix$, while the two H$_2$CO transitions are concentrated more tightly around the continuum structures. The median $\Spix$ remains high over much of the continuum-intensity range, but the three transitions follow different trends and their $\fsim$ values remain well below 100\%. The lowest local similarities occur around parts of the bright continuum structure. High dust intensity does not imply matching molecular-line morphology.

The mismatch toward the continuum peaks is consistent with radiative-transfer effects known in this source. \citet{2025NatAs...9.1833L} derive mean volume densities of order $10^{8}\,{\rm cm}^{-3}$ for the parent disk and $10^{9}\,{\rm cm}^{-3}$ for the compact condensations. \citet{2009ApJ...707....1B} report absorption and self-absorption in the H$_2$CO $3_{03}-2_{02}$ and $3_{22}-2_{21}$ transitions toward compact sources in NGC~6334I(N). Such radiative-transfer effects can alter the observed line morphology relative to the dust continuum. Here the spatial pattern of low $\Spix$ identifies where the observed lines no longer trace the dust structure in the same way.

\subsection{NGC 3627: local, environmental, and scale-dependent tracer correspondence}\label{sec:ngc3627}

NGC~3627 combines resolved maps of molecular gas, ionized gas, and mid-infrared emission over the same galactic disk. For the angular-to-physical scale conversion, we adopt the PHANGS distance of $D=11.32$~Mpc \citep{2021MNRAS.501.3621A}. We use PHANGS--ALMA CO(2--1) \citep{2021ApJS..255...19L,2021ApJS..257...43L}, PHANGS--MUSE H$\beta$, [O~III], H$\alpha$, [N~II], and [S~II] \citep{2022A&A...659A.191E}, and PHANGS--JWST/MIRI F2100W imaging \citep{2023ApJ...944L..17L,2024ApJS..273...13W}. These tracers sample molecular gas, ionized gas, and warm dust associated with recent star formation. We use CO as the reference and ask how the correspondence changes with molecular-gas intensity and spatial scale. The full tracer gallery is shown in Appendix~\ref{app:ngc3627supp}.

\begin{figure*}[t]
\centering
\includegraphics[width=0.96\textwidth]{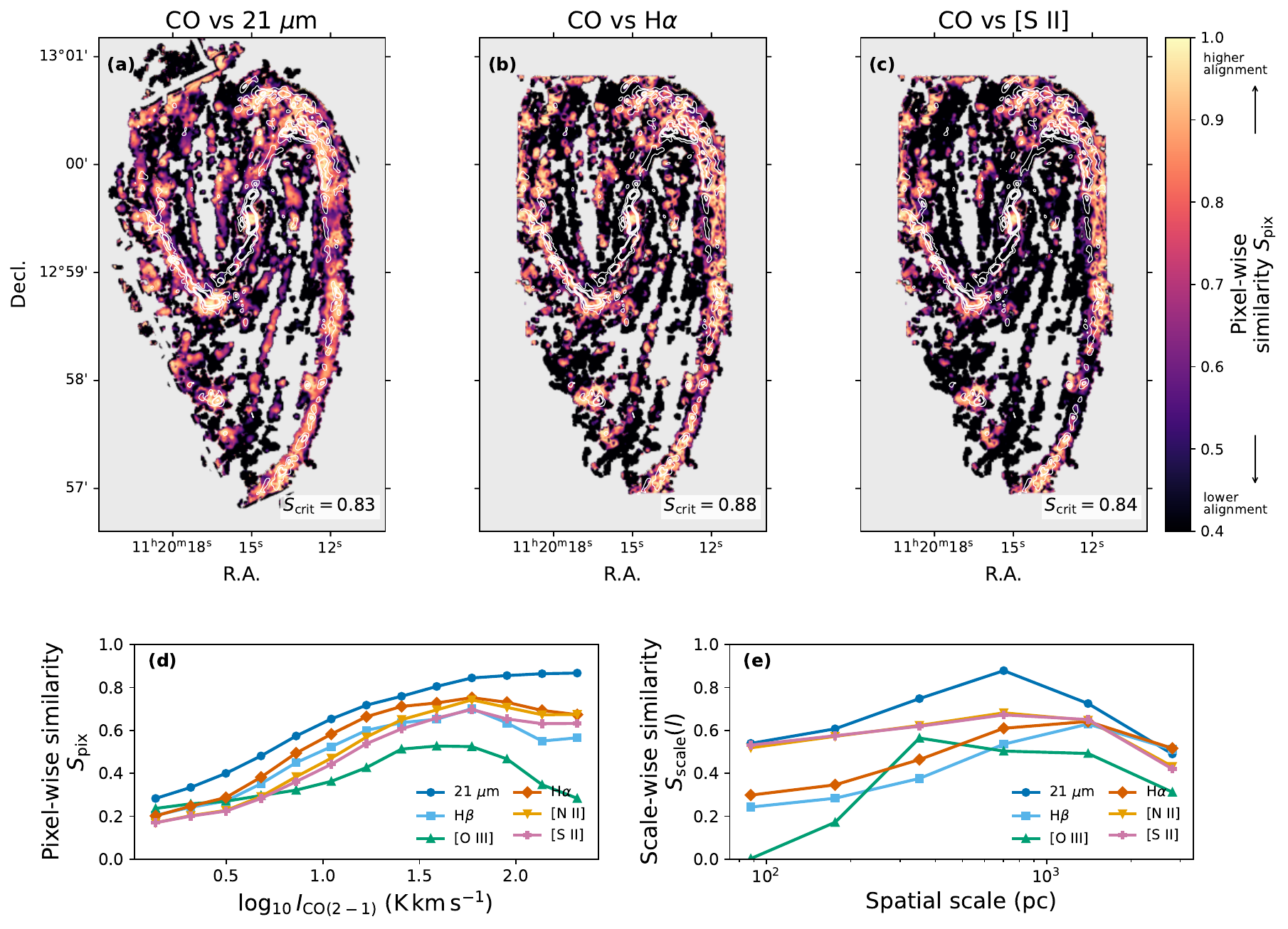}
\caption{Scale-vector similarity in NGC~3627. Top: representative pixel-wise similarity maps for CO(2--1) versus 21~$\mu$m, H$\alpha$, and [S~II]; $\Scrit$ denotes the empirical high-similarity threshold defined by the 95th percentile of the shifted reference distribution. White contours mark the 85th, 95th, and 99th percentiles of the positive CO(2--1) integrated-intensity distribution. Bottom left: median $\Spix$ as a function of CO(2--1) integrated intensity for all six comparison tracers. Bottom right: scale-wise similarity $\Sscale(l)$. The CO--21~$\mu$m comparison reaches the highest scale-wise similarity and peaks at an intermediate sub-kpc scale.}
\label{fig:ngc3627}
\end{figure*}

The local similarity with CO depends on molecular-gas environment. In Figure~\ref{fig:ngc3627}, CO and 21~$\mu$m emission share high-$\Spix$ structure along the bar and spiral features. H$\alpha$ and [S~II] have lower similarity over parts of the disk. Binning by $\ICO$ separates the tracers further. CO--21~$\mu$m similarity rises with CO brightness and remains the highest at the bright end. The optical-line similarities rise more slowly, and H$\beta$ and [O~III] remain lowest. The CO--21~$\mu$m behavior is consistent with the association of 21~$\mu$m emission with embedded star formation and dusty star-forming regions \citep{2023ApJ...944L..21H}.

The same tracer pairs separate in scale space. CO--21~$\mu$m reaches $\Sscale\simeq0.9$ at a scale of order $0.7$~kpc, while the optical lines have lower peaks and different scale dependence. [O~III] has little small-scale correspondence and rises only at intermediate scales. The different scale responses show that molecular gas, dust emission, and ionized gas do not trace identical structures across the disk. This
scale-dependent comparison is related to WWCC, which measures global 
cross-correlation and displacement as a function of spatial scale and can
localize correlated structures at selected scales
\citep{2016A&A...585A..98A}. In SVA, the same CDD representation is used in
two complementary directions: $S_{\rm pix}(x,y)$ compares the full scale
distribution at each position, while $S_{\rm scale}(l)$ compares the spatial
distributions at each scale. The NGC~3627 analysis therefore separates
variations with local environment from variations with spatial scale.

\section{Conclusions}\label{sec:conclusions}

We have introduced scale-vector alignment to compare two intensity fields in CDD scale space. At each position, the CDD components define a scale vector. Its normalized alignment with the corresponding vector of a second tracer gives $\Spix(x,y)$. This normalization removes the overall flux amplitude and retains the relative distribution of intensity over scale. At each spatial scale, the normalized alignment of the corresponding component maps gives $\Sscale(l)$. Spatial shifts define the empirical reference $\Scrit$, from which $\fsim(X)$ can be measured as a function of physical environment.

In Orion~A, the tracer with the highest similarity to the dust-derived column-density map changes from $^{12}$CO to $^{13}$CO to C$^{18}$O as $\NHtwo$ increases. $^{12}$CO has its lowest similarity in OMC-1. In NGC~6334I(N), the line--continuum correspondence decreases around parts of the bright compact structure, where radiative-transfer effects can alter the observed line morphology. In NGC~3627, CO has the highest similarity with 21~$\mu$m emission, and the similarity changes with both CO intensity and spatial scale.

The quantity being compared is the distribution over scale rather than the absolute tracer flux. A change in scale-vector direction marks a change in how a tracer responds to structure of different sizes. Excitation, opacity, chemistry, radiation, and ISM phase can all produce such a change. $\Spix$ shows where it occurs, while $\Sscale$ shows the scales on which the two spatial distributions separate.


\bibliographystyle{aasjournalv7.1}
\bibliography{references}

\appendix

\section{Data preparation and numerical implementation}\label{app:implementation}

Each comparison is performed on a common sky grid and at a common effective angular resolution. For Gaussian beam matching, the higher-resolution map is convolved to the target resolution using
\begin{equation}
{\rm FWHM}_{\rm ker}^2=
{\rm FWHM}_{\rm target}^2-
{\rm FWHM}_{\rm native}^2,
\end{equation}
and then reprojected onto the target WCS. Reprojection footprints and existing masks are propagated through the analysis. The original validity masks are recorded before CDD; masked pixels are set to zero only in the finite numerical arrays passed to the decomposition and are excluded from the reported similarities.

Observed pixel-wise similarities use the common valid footprint of the two input images. In each shift realization, the second-image validity mask is shifted together with its CDD cube, and only the overlap with the original analysis footprint is retained. The normalized inner products are evaluated in double precision after rescaling each local scale vector by its largest absolute component, which avoids overflow without changing the cosine similarity. For $\Sscale$, we retain only CDD channels whose rms amplitude exceeds $10^{-4}$ of the maximum rms amplitude in both images; this prevents numerically negligible residual channels from entering the cosine normalization.

For NGC~6334I(N), the quantitative comparison is restricted to pixels with $I_{\rm cont}\geq5\sigma_{\rm cont}$, where $\sigma_{\rm cont}=1.5\times10^{-4}~{\rm Jy\,beam^{-1}}$. The continuum contours in Figure~\ref{fig:ngc6334} are 20, 50, and $110\sigma_{\rm cont}$. For NGC~3627, the PHANGS--MUSE native-resolution product reports an average PSF FWHM of $0.814''$ in its FITS header. We approximate this PSF as a circular Gaussian when matching the MUSE maps to the PHANGS--ALMA CO(2--1) beam \citep{2022A&A...659A.191E}; the MUSE footprint is then eroded by one target-beam width to remove reprojection-edge structure.

The implementation used in this paper is available as CDD-SVA v0.1.0 at \url{https://github.com/meng-ke/Scale-Vector-Alignment}. It is distributed through PyPI (\url{https://pypi.org/project/cdd-sva/}) and can be installed with \texttt{pip install cdd-sva}. The repository contains the examples used to demonstrate the calculation.

The main calculation can be called as follows:
\begin{verbatim}
from cdd_sva import scale_similarity

result = scale_similarity(
    image_A,
    image_B,
    n_null=200,
    min_shift=30,
)

S_pix   = result.S_pix
S_crit  = result.S_crit
scales  = result.scales
S_scale = result.S_scale
\end{verbatim}
The two input arrays must be on the same pixel grid and at the same effective resolution. Non-finite pixels are excluded from the comparison. The CDD scale sampling and hardware options can also be specified through the package interface.

\section{Synthetic validation}\label{app:synthetic}

\begin{figure*}
\centering
\includegraphics[width=0.96\textwidth]{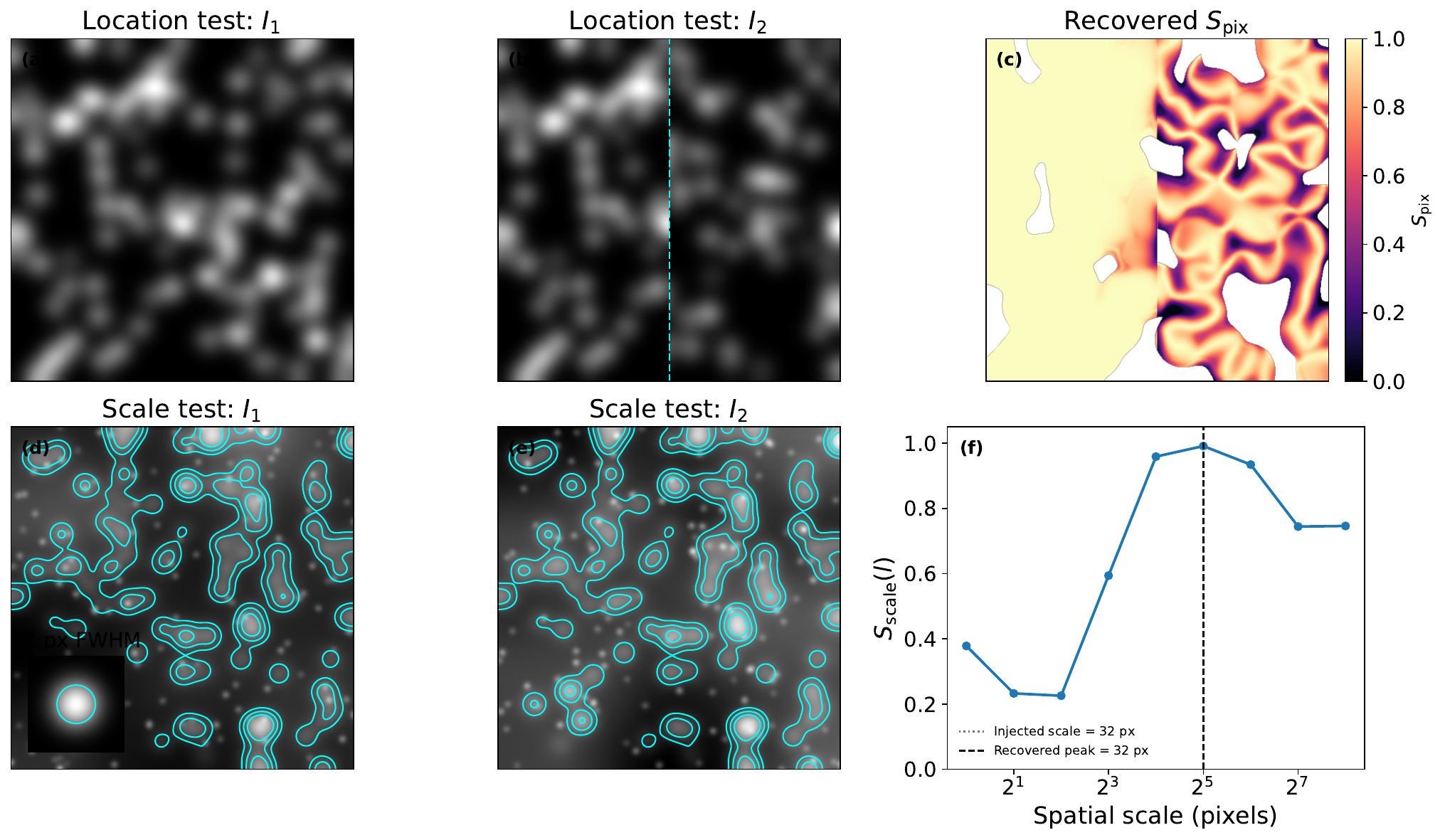}
\caption{Synthetic validation. Top: independent spatial-localization test analogous to the schematic in Figure~\ref{fig:method}. Bottom: scale-recovery test. Cyan contours mark the component shared by the two synthetic images; the inset shows the injected Gaussian kernel with FWHM 32 pixels. The recovered $\Sscale(l)$ peaks at the injected scale.}
\label{fig:synthetic}
\end{figure*}

We test separately the two pieces of information carried by the method: position and scale. The location test in Figure~\ref{fig:synthetic} uses two images that share structure in one region and differ in another. The recovered $\Spix$ follows this imposed spatial correspondence. In the scale-recovery test, the two images contain different small- and large-scale components but share a Gaussian-smoothed component with an injected FWHM of 32 pixels. The recovered $\Sscale(l)$ peaks in the 32-pixel CDD channel. These tests verify that the two statistics recover the imposed correspondence along the dimensions they are designed to measure.

\section{Orion A input and full similarity maps}\label{app:oriondata}

\begin{figure*}
\centering
\includegraphics[width=0.96\textwidth]{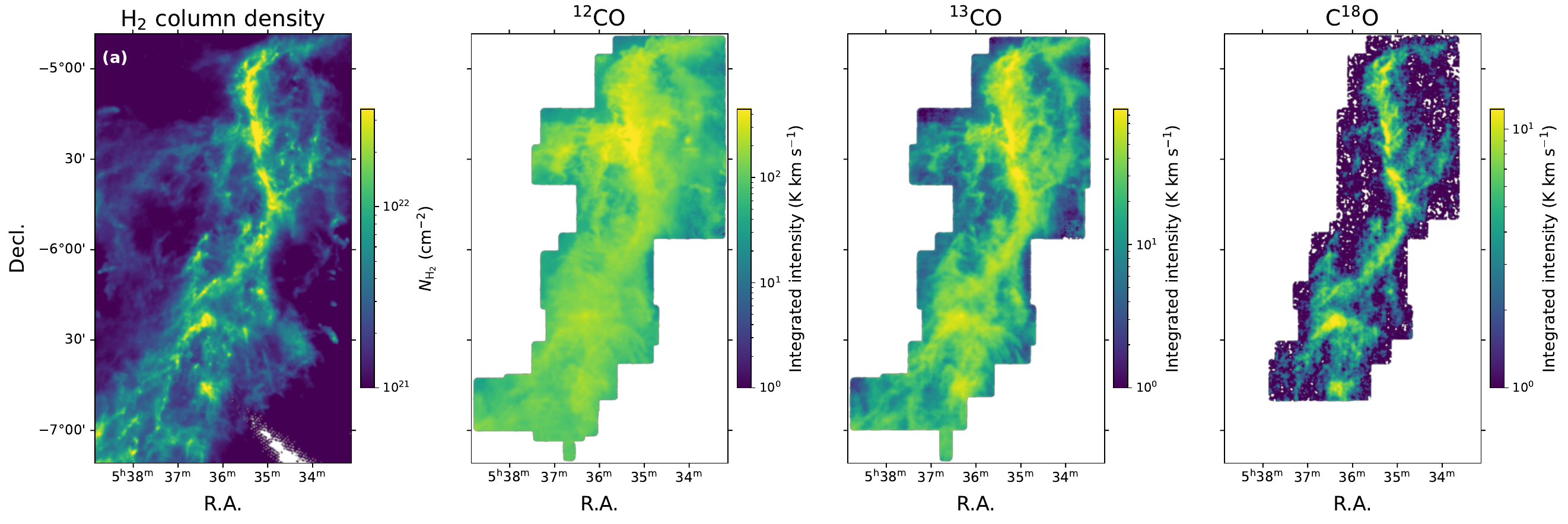}
\caption{Input data for the Orion~A application: H$_2$ column density and $^{12}$CO, $^{13}$CO, and C$^{18}$O integrated intensity.}
\label{fig:oriondata}
\end{figure*}

\begin{figure*}
\centering
\includegraphics[width=0.96\textwidth]{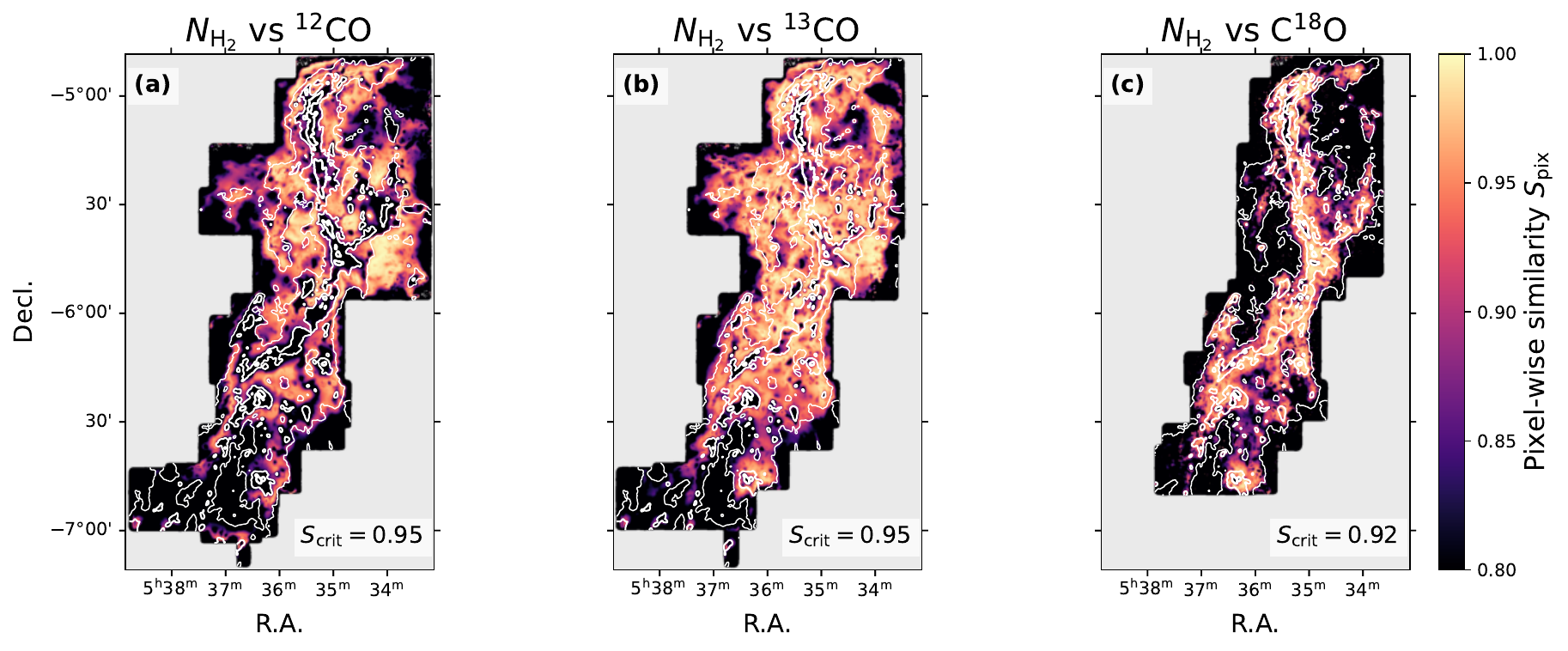}
\caption{Full-field pixel-wise similarity maps for Orion~A. Panels show $\Spix$ between the H$_2$ column-density map and $^{12}$CO, $^{13}$CO, and C$^{18}$O integrated intensity. White contours trace the H$_2$ column-density structure, and each panel gives the tracer-specific empirical threshold $\Scrit$ from the shifted reference distribution. These scalar maps provide the full spatial information summarized by the RGB composite in Figure~\ref{fig:orion}(a).}
\label{fig:orionsimfull}
\end{figure*}

Figure~\ref{fig:oriondata} shows the H$_2$ column-density map and the three CO isotopologue moment-0 maps used in Section~\ref{sec:orion}, after the adopted preprocessing and on their common analysis grid. Figure~\ref{fig:orionsimfull} gives the corresponding full-field scalar $\Spix$ maps that underlie the RGB overview in Figure~\ref{fig:orion}.

\section{NGC 6334I(N) input maps}\label{app:ngc6334data}

\begin{figure*}
\centering
\includegraphics[width=0.96\textwidth]{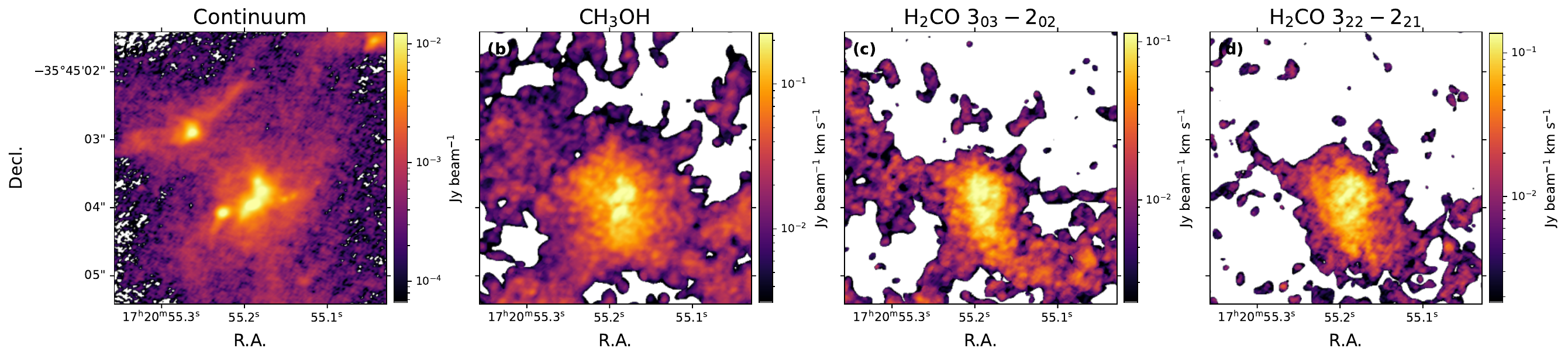}
\caption{Input data for the NGC~6334I(N) application: continuum, CH$_3$OH $4_{2,2}-3_{1,2}$, H$_2$CO $3_{03}-2_{02}$, and H$_2$CO $3_{22}-2_{21}$ maps.}
\label{fig:ngc6334data}
\end{figure*}

Figure~\ref{fig:ngc6334data} shows the continuum and molecular-line products used in Section~\ref{sec:ngc6334}.

\section{NGC 3627 supplementary maps}\label{app:ngc3627supp}

\begin{figure*}
\centering
\includegraphics[width=0.96\textwidth]{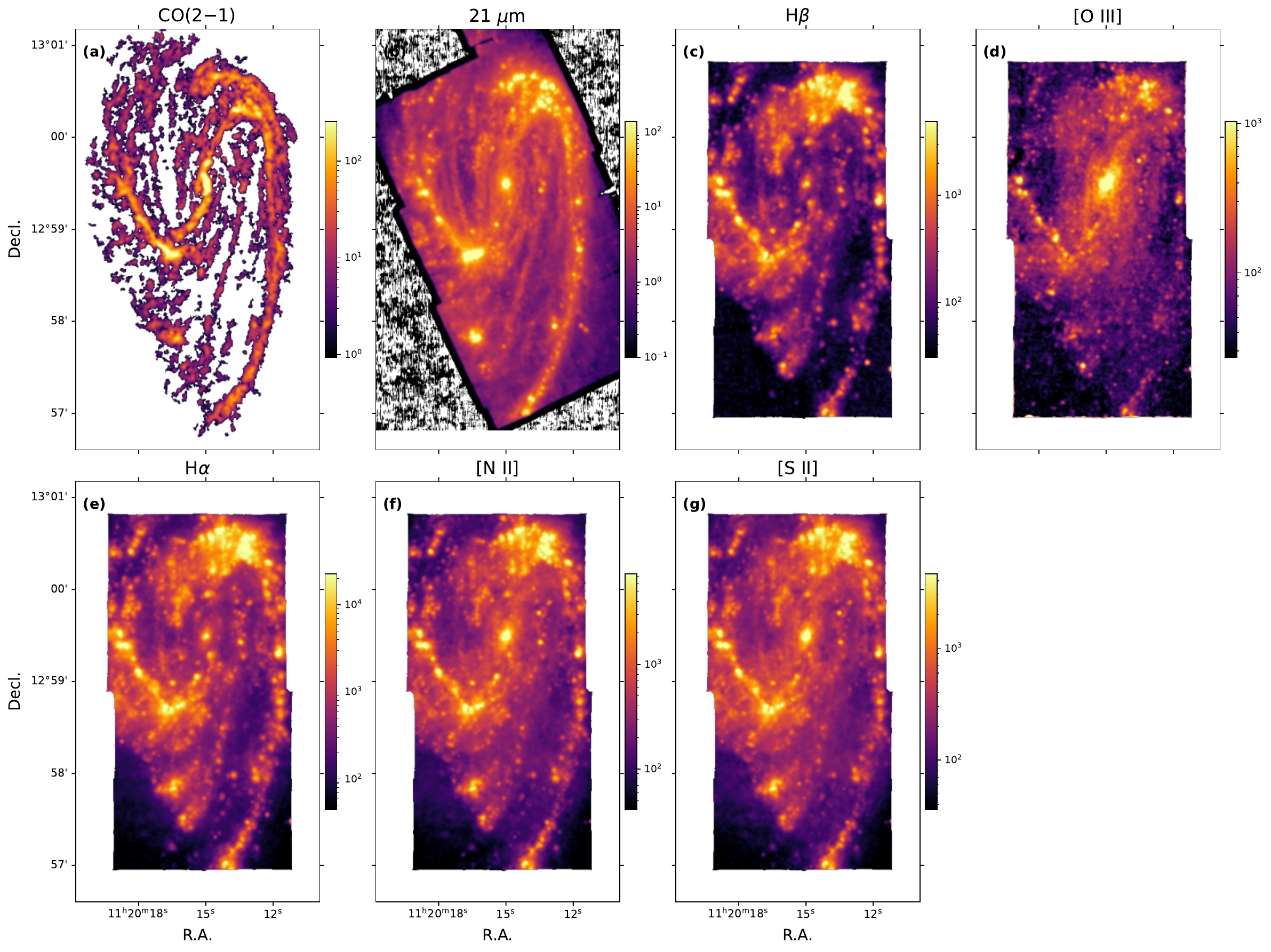}
\caption{Input-data gallery for NGC~3627: PHANGS--ALMA CO(2--1), JWST 21~$\mu$m, and PHANGS--MUSE H$\beta$, [O~III], H$\alpha$, [N~II], and [S~II].}
\label{fig:ngc3627data}
\end{figure*}

\begin{figure*}
\centering
\includegraphics[width=0.80\textwidth]{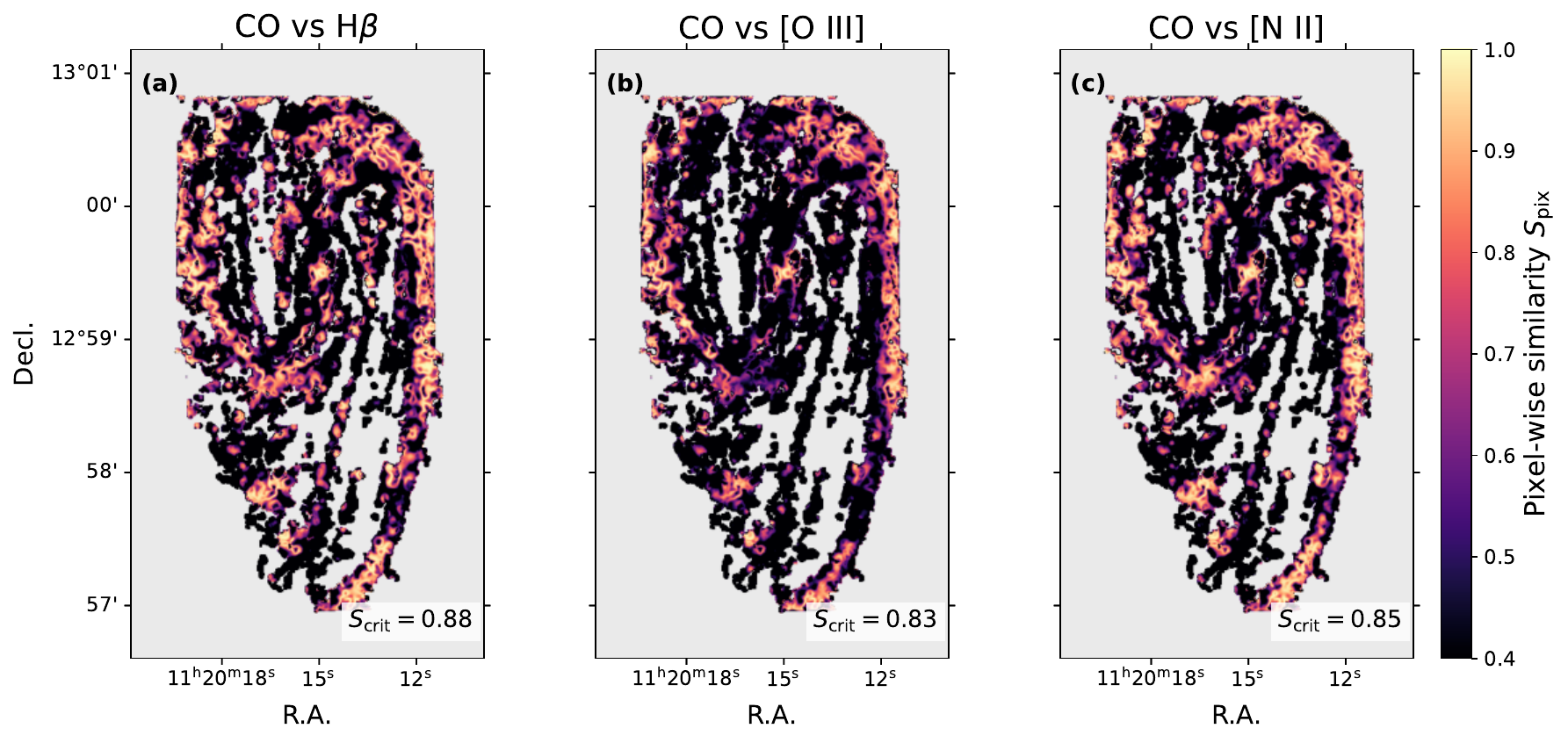}
\caption{Additional NGC~3627 pixel-wise similarity maps for CO versus H$\beta$, [O~III], and [N~II]. The 21~$\mu$m, H$\alpha$, and [S~II] comparisons are shown in Figure~\ref{fig:ngc3627}.}
\label{fig:ngc3627extra}
\end{figure*}

Figure~\ref{fig:ngc3627data} presents the input tracer gallery for NGC~3627. Figure~\ref{fig:ngc3627extra} shows the additional pixel-wise similarity maps omitted from the main figure for visual economy.

\end{document}